\documentclass[runningheads]{llncs}
\usepackage{graphicx, amsmath, amssymb}
\usepackage{xcolor}
\usepackage{trfrac}
\usepackage{algorithm}
\usepackage{algpseudocode}
\usepackage{mathpartir}
\usepackage{multirow}
\usepackage{multicol}
\usepackage{cleveref}
\usepackage{wrapfig}
\usepackage{soul}
\usepackage[title]{appendix}

\newcommand\todo[1]{\textcolor{blue}{TODO: }}

\newcommand\prim[3]{(#1, #2, #3)}
\newcommand{\consexists}{\mathcal{X}^\exists}
\newcommand{\consforall}{\mathcal{X}^\forall}
\newcommand{\tcs}{TC}

\newcommand{\conssym}{\mathcal{X}^{s}}
\newcommand{\timedom}{\mathbb{T}}
\newcommand{\atomicprops}{\mathbb{U}}
\newcommand{\power}{\mathbb{P}}
\newcommand{\globally}[3]{G_{[#1,#2]}(#3)}
\newcommand{\eventually}[3]{F_{[#1,#2]}(#3)}
\newcommand{\flatten}{\textsc{Flat}}
\newcommand{\toolname}{\textsc{StlInc}}
\newcommand{\signal}[0]{\textbf{s}}
\algnewcommand{\IIf}[1]{\State\algorithmicif\ #1\ \algorithmicthen}
\algnewcommand{\EndIIf}{\unskip\ \algorithmicend\ \algorithmicif}

\begin{document}
\title{Safe Planning through Incremental Decomposition of Signal Temporal Logic Specifications}
%
%
\author{Parv Kapoor \inst{1} \and
Eunsuk Kang\inst{1} \and
R\^omulo Meira-G\'oes \inst{2}}
\authorrunning{P. Kapoor et al.}
%
\institute{Carnegie Mellon University, Pittsburgh, PA, USA \and
Pennsylvania State University, State College, PA, USA\\
\email{parvk@cs.cmu.edu, eunsukk@andrew.cmu.edu, romulo@psu.edu}}
\maketitle              
\begin{abstract}
\emph{Trajectory planning} is a critical process that
enables autonomous systems to safely navigate complex environments. \emph{Signal temporal logic (STL)} specifications are an
effective way to encode complex, temporally extended objectives
for trajectory planning in cyber-physical systems (CPS). However,
the complexity of planning with STL using existing techniques scales exponentially with the number of nested operators and the time horizon of a given specification. Additionally, poor performance is
exacerbated at runtime due to limited computational budgets
and compounding modeling errors. Decomposing a complex
specification into smaller subtasks and incrementally planning
for them can remedy these issues. In this work, we present a
method for  decomposing STL specifications  to improve
planning efficiency and performance. The key insight in our
work is to encode all specifications as a set of basic constraints called \emph{reachability}
and \emph{invariance constraints}, and schedule these constraints
sequentially at runtime. Our experiment shows that the
proposed technique outperforms the state-of-the-art trajectory
planning techniques for both linear and non-linear dynamical
systems.

\end{abstract}
\section{Introduction}
\label{sec:intro}
\begin{figure*}
\centering
\begin{tabular}{cccc}
\includegraphics[width=0.95\columnwidth]{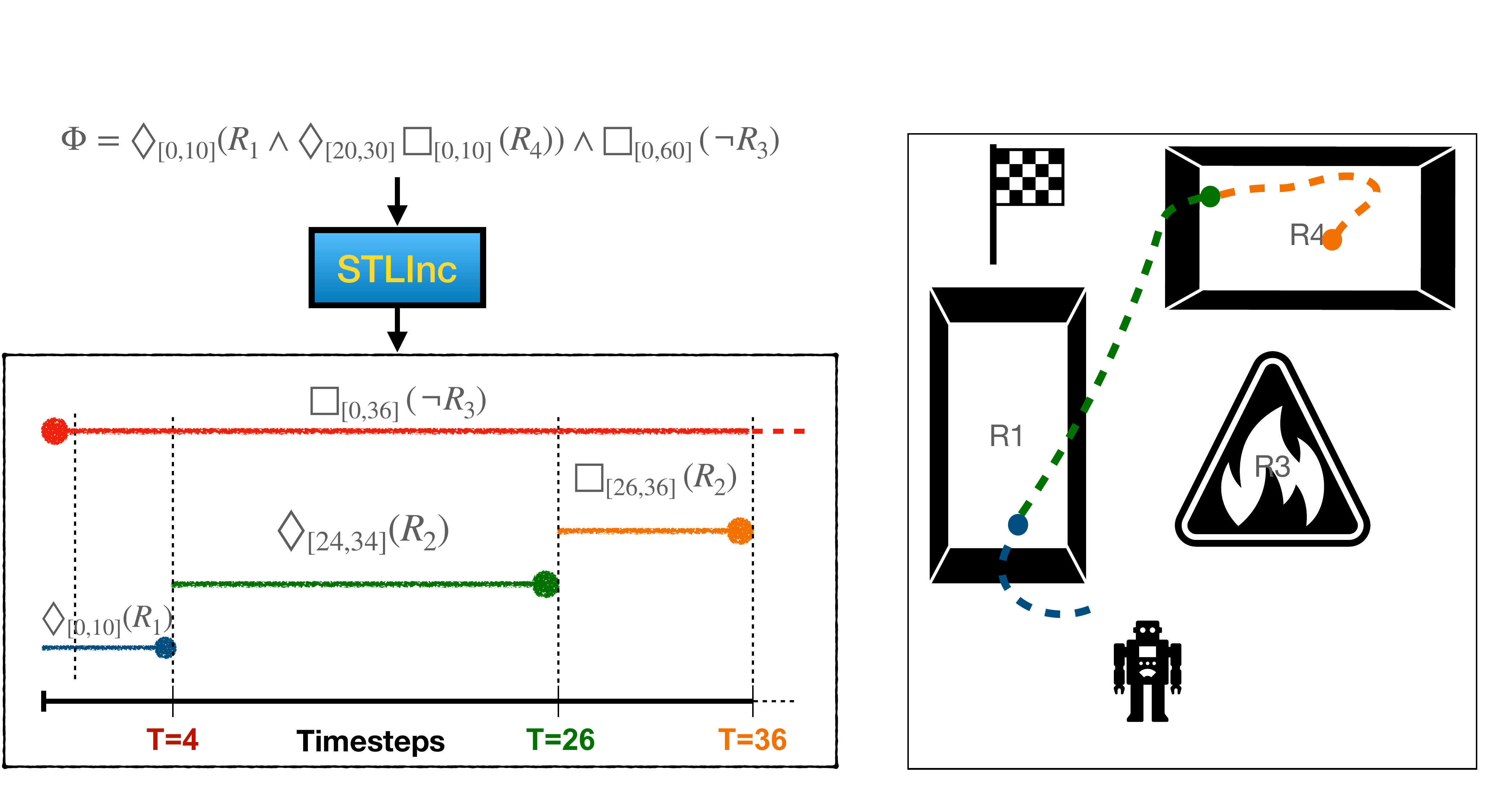}
\end{tabular}
\caption{\textbf{Left}: An STL specification $\phi$ with multiple nested temporal operators and a possible decomposition into subtasks. \textbf{Right}: A sample trajectory that satisfies $\phi$ in a planar environment.} 
\label{fig:pull}
\end{figure*}
Most autonomous robots interacting with the physical world need to achieve complex objectives while dealing with uncertainty and stochasticity in their environment. This problem is exacerbated by  short response times expected while ensuring runtime efficiency. Hence, formulating these complex objectives accurately is a crucial step in realizing the desired behaviors for robotic operations.

Temporal logics such as \emph{linear temporal logic (LTL)}~\cite{4567924} and \emph{signal temporal logic (STL)}~\cite{Maler2004MonitoringTP} provide a precise way to encode objectives that are expressed in a natural language. STL has received special attention in the community due to its rich quantitative semantics that can quantitatively measure satisfaction of a given property that encodes an objective. Additionally, it can be used to describe complex properties over real valued signals such as state trajectories arising from continuous dynamical systems. For robotic planning, STL can be used to describe complex behaviors with concrete time deadlines such as those found in trajectory planning and task planning. Planners can use these specifications to generate  specification-conforming behavior.

A significant amount of common robotic objectives can be interpreted as a sequence of subtasks. It has been shown that incremental subtask planning can be done more efficiently compared to planning for a composite task \cite{10.5555/1661445.1661710,czechowski2021subgoal,Nair2019HierarchicalFS}. 
 However, when STL is used to represent these composite tasks, incremental planning becomes challenging. This issue is because STL semantics can encode the sequential nature of tasks but does not expose this structure to the planner. In such cases, the planners are forced to work with complex long-horizon specifications. When the horizon of the specification is longer than the planning horizon, planners can often generate suboptimal or violating plans. This problem is exacerbated when planning occurs at runtime with computational constraints and compounding modeling errors \cite{10160953}. 

In this work, we propose a theory to decompose long-horizon, arbitrarily nested specifications into sub-specifications that can be satisfied incrementally. We define recursive rules for decomposition and propose a novel scheduling algorithm for incremental task planning. The key insight here is to ``divide and conquer" STL requirements  while ensuring, by construction, that the resulting plan satisfies the original composite specification. We illustrate the effectiveness of our proposed approach over an experiment involving robot exploration problems with linear and non-linear dynamics. Our preliminary experiment shows that our approach is able to more efficiently generate plans for complex, composite specifications in comparison to the existing state-of-the-art STL-based planning methods. In addition, our decomposition technique is agnostic to the underlying system dynamics and the choice of planner, and can potentially be adapted by different planners. 

The key contributions of this paper are: 
\begin{itemize}
    \item A method for decomposing an STL specification into a set of smaller STL specifications that represent subtasks (Section \ref{sec:flattening}); 
    \item A planning algorithm that incrementally schedules and executes these subtasks (Sections \ref{sec:str} and \ref{sec:scheduling}); 
    \item An evaluation of the proposed approach over a benchmark of motion planning tasks (Section \ref{sec:evaluation}).
\end{itemize}

\section{Motivation}
\label{sec:motivation}
We illustrate the problem of planning from complex specifications using an example from the motion planning domain. 
We use a  planning problem similar to the one defined in \cite{leungbackpropagation}.
 
As illustrated in Figure~\ref{fig:pull}, the goal of the agent (robot) is to visit regions $R_{1}$ and $R_{4}$  sequentially while avoiding an unsafe region, $R_3$. Additionally, upon reaching $R_{4}$, the agent needs to stay in it for 10 time steps. We combine three common motion planning patterns such as sequenced visit, stabilization, and global avoidance to create a specification with timed deadlines as follows:
\begin{align*}
        \Phi &= \phi_1 \land \phi_2\\ 
        \phi_1 &= \Diamond_{[0,10]} (R_1 \land \Diamond_{[20,30]} \Box_{[0,10]} (R_4)) \\
        \phi_2 &= \Box_{[0,60]}(\neg R_3)
\end{align*}

The state-of-the-art (SOTA) technique for planning from $\Phi$, originally proposed in \cite{7039363}, involves encoding the STL specification and the system dynamics as Mixed Integer Program (MIP) constraints and solving the constrained optimization problem in a receding horizon fashion. A new binary decision variable is introduced for each atomic proposition per time step in the STL specification. 
A known drawback of this technique is its exponential worst-case complexity with respect to the number of binary variables \cite{kurtz2022mixed}. Various encoding modifications have been suggested to enhance the efficiency of the technique by reducing the number of variables and constraints~\cite{8310901,kurtz2022mixed}. 

However, even with reduced variable encoding, current methods excel primarily with short-horizon specifications. When encoding nested temporal operators, a large number of additional variables and constraints are needed to capture the relationship between different temporal operators, in contrast to non-nested operators, where temporal constraints associated with each operator are considered independently. 
For example, let us take the subformula $\Diamond_{[20,30]} \Box_{[0,10]} (R_4)$ from $\phi_1$. For encoding this subformula into an MIP,  we would need 11 (outer eventually) + 121 (inner always) = 132 binary variables.\footnote{For the outer $\Diamond$ clause, 11 binary variables are introduced to encode that the inner $\Box$ clause is satisfied within interval [20,30]; for each time point in [0,10] interval, another set of 11  variables are introduced, thus resulting in 11*11 = 121 variables.} In general, as the nesting depth increases, the number of variables can increase exponentially.

In this work, we propose a technique to improve the scalability of STL planning algorithms through decomposition of STL specifications. Our idea is inspired by human planning, where long-term goals are achieved by breaking tasks into incremental sub-goals \cite{donnarumma2016problem}. 
Concretely, by decomposing the specification, we effectively remove the complexity of nested operators and also reduce the length of the lookahead horizon. 

A possible decomposition of the specification $\Phi$ into four subtasks is  as follows: 
\begin{align*}
        sch_1 &= \Diamond_{[0,10]} (R1) \land \Box_{[0,10]} (\neg R_3) \\
        sch_2 &=  \Diamond_{[t_{R1}+20, t_{R1} + 30]} (R_4) \land \Box_{[t_{R1}+20,t_{R1} + 30]} ( \neg R_3 )  \\
        sch_3 &=  \Box_{[t_{R4}+0, t_{R4}+ 10]} (\neg R_3 \land R_4) \\
        sch_4 &= \Box_{[t_{R42},60]} (\neg R_3) 
\end{align*}

Here, the symbolic time variables ($t_{R1}, t_{R4}, t_{R42}$) indicate when those subtasks
get satisfied. More specifically, $t_{R4}$ indicates when the agent reaches Region 1 and $t_{R42}$ indicates when the agent has been inside Region 4 for 10 timesteps after reaching it. These variables then shift the time intervals of the other constraints that depend on them (e.g., time $t_{R1}$ from $sch_1$ is used to concretize the time intervals for $sch_2$, whose time of satisfaction, in turn, influences $sch_3$).  
These subtasks have shorter time horizons and no nested temporal operators, resulting in MIP constraints that are less complex than those that would result from composite specifications. As shown later in Section~\ref{sec:evaluation}, this decomposition-based approach has potential to significantly improve the efficiency of planning.

\section{Preliminaries}%
\label{sec:prelims}

STL is a logical formalism used to define properties of continuous time real valued signals \cite{Maler2004MonitoringTP}.
A signal $\mathbf{s}$ is a function $\mathbf{s}:\timedom \to \mathbb{R}^n$ that maps a time domain $T \subseteq \mathbb{R}_{\geq 0}$ to a real valued vector. Then, an STL formula is defined as:
$$\phi := \mu ~|~ \neg \phi ~|~ \phi \land \psi ~|~ \phi \lor \psi ~|~ \phi ~\mathcal{U}_{[a,b]}~\psi$$
where $\mu$ is a predicate on the signal $\mathbf{s}$ at time $t$ in the form of $\mu \equiv \mu(\mathbf{s}(t)) > 0$ and $[a, b]$ is the time interval (or simply $I$). The \emph{until} operator $\mathcal{U}$ defines that $\phi$ must be true until $\psi$ becomes true within a time interval $[a, b]$. 
Two other operators can be derived from \emph{until}: \emph{eventually} ($F_{[a,b]}~\phi := \top~\mathcal{U}_{[a,b]}~\phi$) and \emph{always} ($G_{[a,b]}~\phi := \neg F_{[a,b]}~\neg\phi$).

\begin{definition}
Given a signal $s_{t}$ representing a signal starting at time t, the Boolean semantics of satisfaction of $s_t \models \phi$ are defined inductively as follows:
\end{definition}
\begin{align*}
    s_t\models\mu  &  \iff   \mu(s(t))>0 \\
     s_t \models \lnot \varphi & \iff  \lnot (s_t \models \varphi ) \\
    s_t \models \varphi_1 \land \varphi_2 & \iff  (s_t \models \varphi_1) \land (s_t \models \varphi_2) \\
  s_t \models \text{F}_{[a,b]}(\varphi) & \iff \exists t' \in [t+a, t+b] \text{ s.t. }  s_{t'} \models \varphi \\
     s_{t} \models \text{G}_{[a,b]}(\varphi) &\iff  \forall t' \in [t+a, t+b] \text{ s.t. }  s_{t'} \models \varphi 
\end{align*}
Apart from the Boolean semantics, quantitative semantics are defined for a signal to compute a real-valued metric indicating \textit{robustness}, i.e., the strength of satisfaction or violation. 
\begin{definition} 
Given a signal $s_{t}$ representing a signal starting at time t, the quantitative semantics of satisfaction of $s_t \models \phi$ are defined inductively as follows:
\end{definition}
\begin{align*}
   \rho(s_t, \mu_c) &   =  \mu(x_t) - c  \\
    \rho(s_t,\lnot \varphi) &   = - \rho(s_t, \varphi) \\
    \rho(s_t,\varphi_1 \land \varphi_2) &   = \min( \rho(s_t, \varphi_1), \rho(s_t, \varphi_2))  \\
    \rho(s_t,\text{F}_{[a,b]}(\varphi)) &   = \underset{t' \in [t+a, t+b]}{\max}\rho(s_t', \varphi)  \\
    \rho(s_t,\text{G}_{[a,b]}(\varphi)) &   = \underset{t' \in [t+a, t+b]}{\min}\rho(s_t', \varphi)  
\end{align*}
For example, suppose that we are given (1) $\phi \equiv \text{G}_{[0,3]}(\texttt{distToR3}(t) \geq 3.0)$, which  states that the agent should maintain at least 3.0 meters away from region $R_3$ for the next 4 time steps and (2) signal $s_t$ that contains sequence $\langle 3.0, 2.5, 3.0, 3.5 \rangle$ for $\texttt{distToR3}$. Evaluting the robustness of satisfaction of $\phi$ over $s_t$ would result in a value of $-0.5$, implying that the agent violates the property by a degree of $0.5$ (i.e., it fails to stay away from $R_3$ by 0.5 meters).
\section{Approach}%
\label{sec:controller-synthesis}

\begin{figure*}[!t]
     \centering  
    \includegraphics[width=0.9\textwidth]{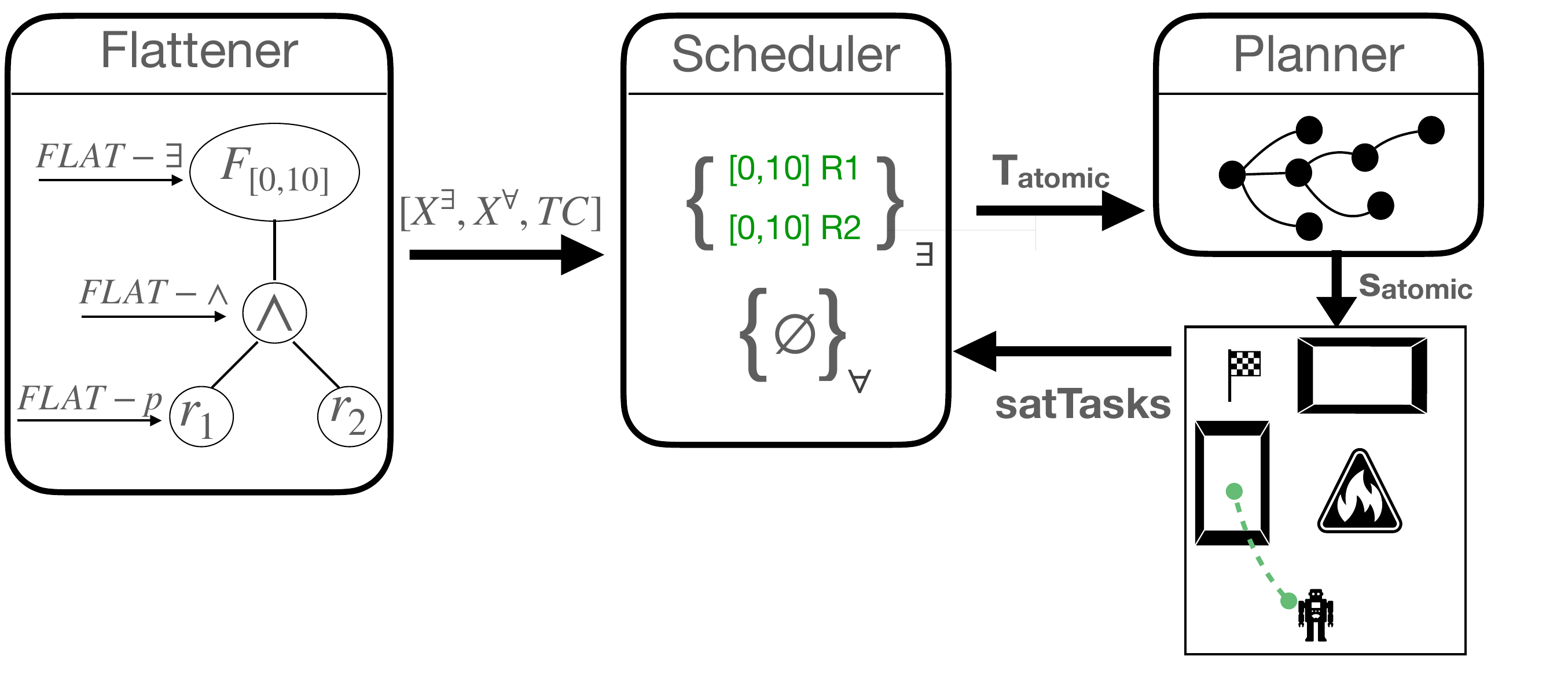}    
      \caption{Overview of the \toolname{} approach}
     \label{fig:goalrep}
\end{figure*}
\subsection{Basic Concepts and Definitions} \label{sec:basic}

The overview of our planning framework (\toolname{}) is shown in Figure~\ref{fig:goalrep}. The key idea behind this approach is that a bounded STL formula in our fragment can be decomposed into a finite set of the following two types of \emph{task constraints}, each of which is associated with time interval $I=[a,b]$ and state proposition $p$:
\begin{quote}
\emph{Reachability}: The system ensures that $p$ holds over at least one time step $t$ within I.\\
\emph{Invariance}: The system ensures that $p$ holds over every  step $t$ within $I$.
\end{quote}
Based on this idea, the framework carries out the incremental planning process over three steps. First, the \emph{flattener} takes a user-specified  STL specification ($\phi$)  and decomposes it into two sets of task constraints, $\consexists$ and $\consforall$, which contain the reachability and invariance constraints, respectively. The decomposition is performed such that satisfying all of the constraints in these two sets implies the satisfaction of the original formula $\phi$. 

Next, the \emph{scheduler} takes the two sets, $\consexists$ and $\consforall$, and generates a sequence of \emph{atomic tasks}, $\sigma = \langle at_0, at_1...at_k\rangle$, where (1) each atomic task $at$ is a non-nested STL formula consisting of $\globally{a}{b}{p}$ or $\eventually{a}{b}{p}$ (where $p$ is a propositional formula) and (2) two different atomic tasks do not overlap in their time intervals. 
After this sequence is generated, the \emph{planner} executes these atomic tasks one by one in the designated order. Once all tasks are executed, the system will have fulfilled $\consexists$ and $\consforall$, thus satisfying the original goal of $\phi$. The rest of this section describes each of the three steps in detail.

\paragraph{\textbf{STL fragment}} Our approach is designed to handle specifications written in the following fragment of STL: 
\begin{align*}
    \phi &::= F_{[a,b]} (\phi) \mid G_{[a,b]} (\phi) \mid \phi_1 \land \phi_2 \mid p \\
    p &::= p_1 \land p_2 \mid p_1 \lor p_2 \mid \neg p \mid \mu
\end{align*}
where $p$ is a propositional formula that does not contain any temporal operator, and $\mu$ is an atomic proposition. We target this STL fragment as (1) it signficantly simplifies the decomposition process and (2) it is still expressive enough to capture many common behavioral patterns in the robotics planning domain~\cite{8859226}. Note that the fragment allows a nesting of temporal operators of an arbitrary depth, which is important for specifying sequential tasks. For example, the objective of \emph{visiting}  locations in a particular order can be defined as: 
\begin{align*}
    \phi \equiv F_{[1,5]} (l_1 \land F_{[1,10]} (l_2 \land F_{[1,7]}(l_3))) 
\end{align*}
where $l_1, l_2, l_3$ represent the locations to be visited. 

\paragraph{\textbf{Task constraints}} As mentioned at the beginning of the section, an STL formula imposes two types of constraints over system behavior: \emph{reachability} and \emph{invariance} constraints, which are formally defined as follows:
\begin{definition}[Reachability set]
A reachability set, $\consexists \in \power(\timedom \times \timedom \times \atomicprops)$ is a set of tuples of form $\prim{l}{h}{p}$, each stating that there exists some time  $t$ in interval $[l, h]$, inclusively, such that proposition $p$ holds over the system state at time $t$.
\end{definition}
\begin{definition}[Invariance set]
An invariance set, $\consforall \in \power(\timedom \times \timedom \times \atomicprops)$, is a set of tuples of form $\prim{l}{h}{p}$, each stating that for every time $t$ in interval $[l, h]$, inclusively, proposition $p$ holds over the system state at time $t$.
\end{definition}

We introduce an additional set of constraints that specify intervals over symbolic time variables that are introduced during the flattening of $F$ formulas:
\begin{definition}[Time variable intervals]
A time variable interval set, $\tcs \in \power (\timedom \times \timedom \times \timedom)$, is a set of tuples of the form \prim{l}{h}{v}, each stating that symbolic time variable $v$ takes a  value in the interval $[l, h]$, inclusively.
\end{definition}

\subsection{Flattening}
\label{sec:flattening}

Given an input STL specification, $\phi$, the goal of \emph{flattening} is to construct two sets of constraints---reachability and invariance sets---whose satisfaction also implies the satisfaction of $\phi$. 
Flattening is applied recursively based on the structure of $\phi$, as shown through the rules in Figure~\ref{fig:flattening-rules}. Along with $\consforall$ and $\consexists$, each recursive step produces an additional auxiliary output $\tcs$, which is later used by the scheduler to resolve symbolic time variables.

\begin{figure*}[!t]
\centering
{\small
\begin{tabular}{cc}
$
\trfrac[\ \flatten-$p$]
{\phi = p}
{
\begin{tralign}
\consexists & = \{ (0, 0, p)  \} \quad \consforall = \emptyset \\
\tcs &= \emptyset \\
\end{tralign}
}
$
\qquad
&
$
\trfrac[\ \flatten-$\land$]
{ 
\begin{trgather}

\phi = \phi_{1} \land \phi_{2} \\ 
(\consexists_{1}, \consforall_{1}, \tcs_1) =\flatten(\phi_{1}) \\
(\consexists_{2}, \consforall_{2}, \tcs_2) =\flatten(\phi_{2}) 
\end{trgather}
}
{
\begin{tralign}
\consexists & = \consexists_{1} \cup \consexists_{2}\\ 
\consforall & = \consforall_{1} \cup \consforall_{2} \\
\tcs & = \tcs_1 \cup \tcs_2  \\
\end{tralign}}
$
\\
 & 
\\
\multicolumn{2}{c}{
$
\trfrac[\flatten-$\exists$]
{
\begin{trgather}
\phi = F_{[a,b]}\phi_1 \qquad 
(\consexists_1, \consforall_1, \tcs_1) =\flatten(\phi_1)
\end{trgather}
}
{
\begin{tralign}
 \consexists & = \{ \prim{t+l}{t+h}{p}\ |\  \exists \prim{l}{h}{p} \in  \consexists_1\}\ \\
 \consforall & = \{\prim{t + l}{t + h}{p}\ |\ \exists \prim{l}{h}{p} \in \consforall_1 \} \\
\tcs & = \tcs_1 \cup \{(a,b, t)\} \\

\end{tralign}
}
$}
\\
 & 
\\
\multicolumn{2}{c}{
$
\trfrac[\flatten-$\forall$]
{
\begin{trgather}
\phi = G_{[a,b]}\phi_1, \ \
(\consexists_1, \consforall_1, \tcs_1) =\flatten(\phi_1)
\end{trgather}
}
{
\begin{tralign}
  \consexists & = \{ (k + l, k + l, p) \ |\ \exists (l, h, p) \in \consexists_1 \land k \in [a, b]\} \\
\consforall & = \{ \prim{a+l}{b+h}{p}\ |\  \exists \prim{l}{h}{p} \in  \consforall_{1} \} \\
\tcs & = \tcs_1  \\
\end{tralign}
}
$
}
\\
\end{tabular}
}
\caption{Flattening rules}
\label{fig:flattening-rules}
\end{figure*}

\subsubsection{\flatten-$p$}

In the basic case where $\phi = p$, flattening generates one reachability constraint that requires $p$ to be satisfied at the current time (i.e., $l = h = 0$). Hence, the reachability set for $\phi$ is $(0,0,p)$.

\subsubsection{\flatten-$\land$}

Given cojunctive formula $\phi = \phi_1 \land \phi_2$, the invariance set for $\phi$ is the union of $\consforall_1$ and $\consforall_2$; i.e., every invariance constraint in $\phi_1$ and $\phi_2$ must be satisfied. Similarly, the reachability set for $\phi$ is the union of $\consexists_1$ and $\consexists_2$; i.e., every reachability constraint in $\phi_1$ and $\phi_2$ must be satisfied. 

\subsubsection{\flatten-$\exists$\label{sec:flattenex}}
Given $\phi = \eventually{a}{b}{\phi_1}$, for each constraint $(l_1, h_1, p_1) \in \consexists_1$, flattening involves shifting interval $[l_1,h_1]$ by a symbolic time variable $t \in [a,b]$. This bound on $t$ is encoded by adding a tuple $(a,b,t)$ to $\tcs$. Intuitively, this can be understood from the following transformation similar to the one  in  \cite{10.1007/978-3-031-33170-1_14}:
\begin{align*}
    F_{[a,b]}(F_{[c,d]} (\phi_1)) = F_{[a+c, b+d]}(\phi_1)
\end{align*}
Here, instead of adding $a$ and $b$ directly to the upper and lower bound, respectively of  time interval $[c,d]$, we add a symbolic variable that can take a value between $a$ and $b$. For example, given specification $\phi = \eventually{2}{5}{p}$, $\consexists$ would contain constraint $(t + 0, t + 0, p)$ and $\tcs$ would contain $(2, 5,t)$.

As another example, consider specification
 $\phi = \eventually{2}{5}{p_2 \land \eventually{1}{3}{p_1}}$. When the \flatten-$\exists$ rule is applied, the resulting $\consexists$ contains $(t_2 + t_1 + 0,t_2 + t_1 + 0, p1)$ and $(t_2,t_2,p2)$, and $\tcs$ contains $\{(1, 3, t_1)$, $(2, 5, t_2)\}$.
Here, $t_1$ is introduced when \flatten-$\exists$ rule is applied to the innermost F operator ($F_{[1,3]}$); 
$t_2$ is then introduced when \flatten-$\exists$ is applied for outermost F operator ($F_{[2,5]}$). 

The flattening of an $F$ formula applied to an invariance set is handled similarly. Consider a specification of the form
\begin{equation*}
\phi \equiv  F_{[a,b]}(\phi_1) = F_{[a,b]}(G_{[c,d]} (p)) \label{reachstay}
\end{equation*}
with $\consforall_1 = \{\prim{c}{d}{p}\}$. Here, according to the   Boolean STL semantics, there exists a $t \in [a,b]$ such that $\forall t' \in [t+c,t+d]$, p must be satisfied. Hence, in the resulting invariance set for $\phi$, the time interval for the existing invariance constraint is shifted by $t$. This bound on t is encoded by adding a tuple $(a,b,t)$ to $\tcs$. For example, given specifcation $\phi = F_{[2,5]}(G_{[3,10]} (p)) $, $\consforall$ would contain constraint $(t + 3, t + 10, p)$ and $\tcs$ would contain $(2, 5, t)$. 

\subsubsection{\flatten-$\forall$}
The flattening of an $\forall$ formula over $\consexists_{1}$ is handled in the following way. Consider a specification of the form:
\begin{equation*}
\phi \equiv  G_{[a,b]}(\phi_1) = G_{[a,b]}(F_{[c,d]} (p)) \label{alwayseventually}
\end{equation*}
with $\consexists_1 = \{tc\} = \{\prim{t_1}{t_1}{p}\}$ and $\tcs = \{ (c,d, t_1)\}$. Since $\phi$ states that constraint $tc$ must hold at every time step between $[a,b]$, the idea is to create multiple reachability constraints of form $(t_1 + k, t_1 + k, p)$, one for each time value $k$ in interval $[a,b]$.  For example, let $\phi = \globally{1}{100}{\eventually{1}{5}{p}}$. After flattening, $\consexists$ for $\phi$ would contain 100 reachability constraints, each in form of $(t_1 + k, t_1 + k, p)$ for $1 \leq k \leq 100$.

The flattening of an $\forall$ over $\consforall_{1}$ is handled in the following way. Consider a specification of the form
\begin{equation*}
\phi \equiv  G_{[a,b]}(\phi_1) = G_{[a,b]}(G_{[c,d]} (p)) \label{always2}
\end{equation*}
 with $\consforall_1 = \{tc\} = \{\prim{c}{d}{p}\}$. Since $\phi$ states that constraint $tc$ must hold at every time step between $[a,b]$, by shifting the interval of each constraint by $[a,b]$, the resulting formula  $\consforall$ is constructed as $\{\prim{a+c}{b+ d}{p}\}$.  Intuitively, this can be understood from the following transformation similar to the one in  \cite{10.1007/978-3-031-33170-1_14}:
\begin{align*}
    G_{[a,b]}(G_{[c,d]} (\phi_1)) = G_{[a+c, b+d]}(\phi_1)
\end{align*}

We now introduce a theorem that establishes the soundness of our flattening operation with respect to a given STL specification. 
\begin{lemma}
Let $\phi$ be an input STL specification, and let ($\consexists$, $\consforall$, $TC$) be the output of $flatten(\phi)$. Then, for every  signal $s_t$:
\[
\forall v \in \mathcal{V}(vars(TC)) \bullet \forall x \in \consexists \cup \consforall \bullet s_t \models Inst(x,v) \implies  s_t \models \phi 
\]
where $vars$ is a function that returns the set of all symbolic variables in $\tcs$, $\mathcal{V(X)}$ is the set of all possible assignments of values to variables in $\mathcal{X}$ (restricted to values in their respective time intervals), and $Inst(x, v)$ instantiates symbolic variables in constraint $x$ with the values from $v$.   
\end{lemma}
In other words, if every possible reachability or invariance constraint is satisfied over a particular signal, then the original specification $\phi$ must also hold over the same signal.  

\subsection{Symbolic Time Resolution}
\label{sec:str}

The $\consexists$ and $\consforall$ constraints generated from flattening may contain multiple symbolic variables. To enable scheduling, \toolname{} first attempts to resolve as many of these variables as possible by applying substitution rules to the constraints. 

Algorithm~\ref{alg:PSSR} shows the sketch of the \emph{symbolic time resolution (STR)} process. 
As inputs, it takes the output of the flattening procedure---the $\consexists$ and $\consforall$ constraints, and the time intervals over symbolic time variables. As  outputs, it produces (1) a new pair of $\consexists$ and $\consforall$, with some of the time variables replaced by concrete time values, and (2)  time variable intervals ($\tcs'$) that specifies, for each concrete constraint produced in (1), an interval during which the constraint must be satisfied. This latter set of time intervals are used to enforce conjunctive constraints (i.e., $\phi_1 \land \phi_2$) to be satisfied simultaneously.

The algorithm iterates through the time variables in the bottom-up order (i.e., starting with the variables that appear in the lower part of an AST for a given STL expression). For each constraint that contains variable $t$ (line 4), \textsc{STR} applies two types of substitutions, depending on whether the constraint is a reachability or an invariance constraint.

\begin{algorithm}[!t]
  \caption{Symbolic Time Resolution (\textsc{STR})
  }\label{alg:PSSR}
  \begin{algorithmic}[1]
    \State \textbf{Input}: Flattened constraints $\consexists, \consforall$, time variable intervals $\tcs$
    \State \textbf{Output}: Modified constraints $\consexists, \consforall$, new time variable intervals $\tcs'$
    \State $\tcs' = \emptyset$
    \For{$tc=(a, b, t)$ in $\tcs$} 
        \State $\consforall_t$ := \textsc{filter}($\consforall, t$), $\consexists_t$ := \textsc{filter}($\consexists, t$)   
        \For{$x=(l,h,p)$ in $\consforall_t$}
            \State ($y^\exists, y^\forall$) := \textsc{ApplyFG}($x$,$tc$)
            \State $\consexists := \consexists \cup \{y^\exists\}$
            \State $\consforall := (\consforall - \{x\}) \cup \{y^\forall\}$
            \State $\tcs' := \tcs' \cup \{ (t+l, t+l, \textsc{t-sat}(y^\exists)), (t + l + 1, t + h, \textsc{t-sat}(y^\forall))  \}$
        \EndFor
        \For{$x=(l,h,p)$ in $\consexists_t$}
            \State $y^\exists$ := \textsc{ApplyFF}($x$,$tc$)
            \State $\consexists := (\consexists - \{x\})\cup \{y^\exists\}$
            \State $\tcs' := \tcs' \cup \{ (t + l, t + h, \textsc{t-sat}(y^\exists)) \}$
        \EndFor
    \EndFor
    \State \textbf{return} ($\consexists, \consforall, \tcs'$)
  \end{algorithmic}
\end{algorithm}

\subsubsection{Invariance constraints}

An invariance constraint, $x = (l, h, p)$, with time variable $t \in [a,b]$ represents an STL expression $F_{[a, b]}(G_{[l,h]}\ p)$ (where $t$ appears in $l$ and $h$, and $p$ is an atomic proposition). Intuitively, this expression can be regarded a kind of ``reach and stay'' task, where the system must first reach a state where $p$ holds within time interval $[a + l, b + l]$, and then continue to satisfy $p$ for the following $(h-l)$ time steps. The  rule $\textsc{ApplyFG}$ takes constraint $x$ and interval $tc$, and produces a new pair of reachability and invariance constraints, as follows:
\begin{align*}
\textsc{ApplyFG}(x=(l, h, p),tc=(a,b,t))
\equiv ((a + l, b+l, p), (t_{sat}, t_{sat} + h-l,p))
\end{align*}
where $t_{sat}$ is a new symbolic variable that represents the time at which $p$ is satisfied between $(a + l, b + l)$.

In addition, \textsc{STR} adds two time variable intervals (line 10) to ensure that: (1) the new  reachability constraint, $y^\exists$, is satisfied exactly at $(t + l)$ and (2) the invariance constraint, $y^\forall$, is satisfied subsequently for the following $(l - h)$ steps. Here, $\textsc{t-sat}(y)$ returns a symbolic time variable representing the time of satisfaction of constraint $y$.

Consider the example in Figure~\ref{fig:tree}. One of the invariance constraints that flattening generates is $(t+1, t+5, r_1)$, which depends on time variable $t \in [1,20]$. The application of $\textsc{ApplyFG}$ results in constraints stating that (1) the system must satisfy $p$ within [2,21], and (2) from the point of the satisfaction of this constraint ($t_2$), it must hold $p$ true for the following $(5-1) = 4$ steps. Note that the other invariance constraint, $(1, 35, \neg r_3)$, remains untouched, as it does not depend on any time variable.
\subsubsection{Reachability constraints}
A reachability constraint, $x = (l, h, p)$, with time variable $t \in [a,b]$ represents an STL expression $F_{[a, b]}(F_{[l,h]}\ p)$. When a pair of $F$ operators are nested in this manner, they can be simplified by using the following substitution rule:
\begin{align*}
\textsc{ApplyFF}(x=(l, h, p),tc=(a,b,t))
\equiv (l + a, h + b, p)
\end{align*}
The resulting reachability constraint, $y^\exists$, replaces the existing constraint in $\consexists$ (line 14). In addition, a new time variable interval is added to ensure that $y^\exists$ is satisfied between $(t + l)$ and $(t + h)$.

\begin{figure*}[!t]
     \centering  
    \includegraphics[scale=0.30]{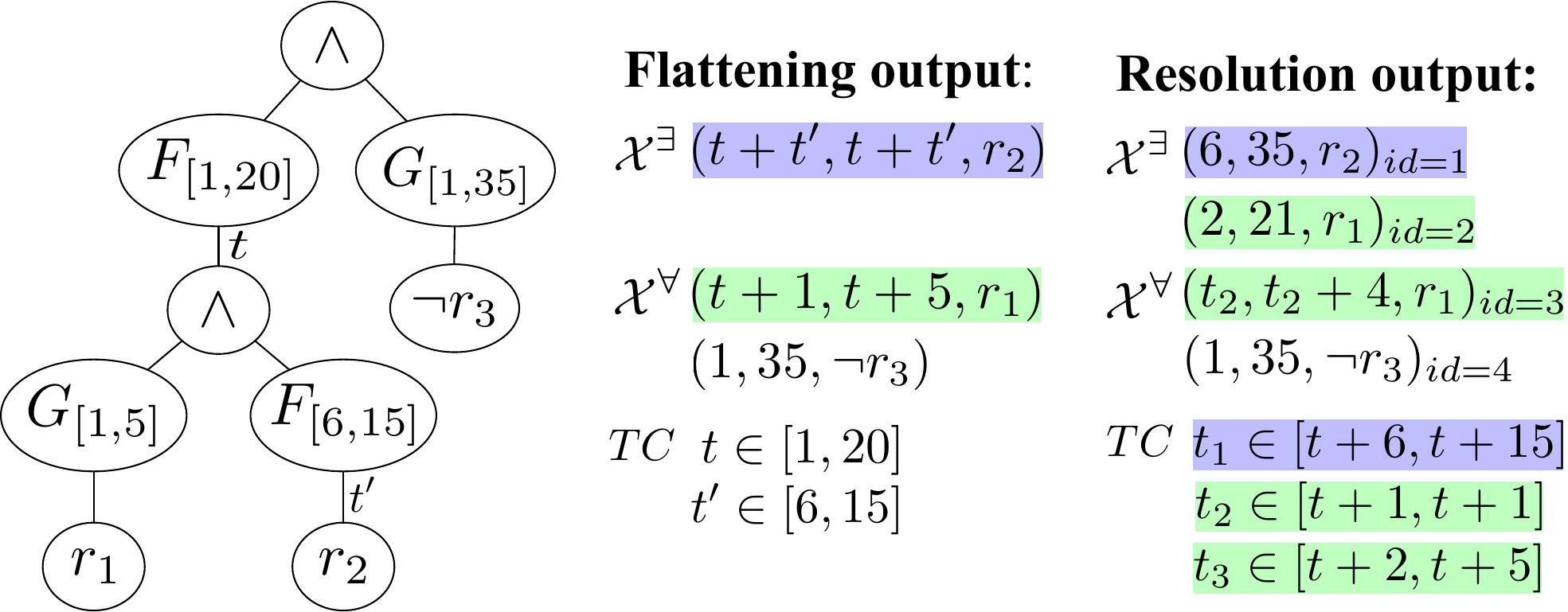}    
      \caption{An AST for an example STL specification, $F_{[1,20]}(G_{[1,5]} (r_1) \land F_{[6,15]} (r_2)) \land G_{[1,35]} (\neg r_3)$, with outputs from the flattening and symbolic time resolution steps. Each constraint resulting from the resolution step is assigned an identifier ($id$); a symbolic time variable that represents the time of satisfying the constraint is annotated with $id$ as the subscript (i.e., $t_2$ represents the satisfaction of reachability constraint $(2, 21, r_1)$).
      } \label{fig:tree}
\end{figure*}

For the example in  Figure~\ref{fig:tree}, flattening produces one reachability constraint, $(t + t', t + t', r_2)$. During the first iteration of the outermost loop (line 4),  \textsc{STR}  selects $tc = (5, 15, t')$ and applies $\textsc{ApplyFF}$ to the constraint, producing a new reachability constraint $(t + 5, t + 15, r_2)$. In next iteration, \textsc{STR} selects $tc = (1, 20, t)$ and applies $\textsc{ApplyFF}$ to $(t + 5, t + 15, r_2)$, producing an additional reachability constraint $(6, 35, r_2)$. 

Note that at the end of resolution for the example (Figure~\ref{fig:tree}), variable $t$ appears in multiple intervals in $\tcs'$. In the following section, we describe how $\toolname{}$ schedules tasks to generate concrete values for the time variables incrementally one-by-one, ultimately synthesizing a plan that satisfies the original STL formula. 

\begin{lemma}
Let ($\consexists$, $\consforall$, $TC$) be the output of $flatten(\phi)$ and ($\consexists_{2}$, $\consforall_{2}$, $TC'$) be the output of $STR (\consexists, \consforall, TC)$.  Then, for every signal $s_t$:
\begin{align*}
\forall v \in \mathcal{V}(vars(TC')) \bullet (\forall x_2 \in \consexists_2 \cup \consforall_2 \bullet s_t \models Inst(x_2,v)) \implies  \\ 
(\forall x \in \consexists \cup \consforall \bullet s_t \models Inst(x,v))
\end{align*}
\end{lemma}
In other words, if every reachability and invariance constraint generated from STR is satisfied under some instantiation ($v$) of symbolic time variables in $TC'$, then the original set of flattened constraints must also be satisfied under the same condition.

\subsection{Scheduling}
\label{sec:scheduling}
\begin{algorithm}[!t]
  \caption{\label{alg:schedule} Schedule}
  \begin{algorithmic}[1]
    \State \textbf{Input}: $\consexists, \consforall, \tcs'$
    \State \textbf{Output}: Signal $\signal_{plan}$ for 
    synthesized plan
    \State $\signal_{plan}= \langle \rangle$
    \State $\prec$ := \textsc{ComputeOrder}($\consexists, \consforall, \tcs'$)
    \State $currTasks$ := \textsc{NextTasks}($\consexists, \consforall, \prec, \emptyset$)
    \While{$currTasks \neq \emptyset$}
        \State $currTasks$ := \textsc{Slice}($currTasks$)
        \State $atomicTask$ := \textsc{NextAtomic}($currTasks$)
        \State $currTasks$ := $currTasks - \{atomicTask\}$
        \State $\signal_{atomic}$ := \textsc{Plan}
        ($atomicTask$)
        \IIf{$\signal_{atomic} = \langle \rangle$} \textbf{break} \EndIIf 
        \State $\signal_{plan} := \signal^\frown\signal_{atomic}$
        \State satTasks := \textsc{ExtractTime($\signal_{atomic}$)}
        \State $currTasks$ := $currTasks \cup \textsc{NextTasks}(\consexists, \consforall, \prec, satTasks)$
    \EndWhile
    \State \textbf{return} $\signal_{plan}$
  \end{algorithmic}
\end{algorithm}

Given the output from the resolution step (the two constraint sets, $\consexists, \consforall$ and the time interval variables, $\tcs'$), the goal of the scheduler is synthesize a plan that satisfies the original STL specification $\phi$. To achieve this, the scheduler iteratively interacts with a \emph{planner} that is capable of synthesizing a plan to satisfy an \emph{atomic task} formula of form $F_{[a,b]}(\phi_1) \land G_{[a,b]}(\phi_2)$, where $\phi_1$ and $\phi_2$ are quantifier-free STL expressions. The scheduling process is incremental: The scheduler generates a sequence of atomic tasks formulas and invokes the planner to solve them one-by-one, using information (i.e., the time of satisfaction of a reachability or invariance constraint) generated by the planner to resolve any dependencies on symbolic time variables that were introduced during the flattening and resolution steps. The scheduling algorithm (Alg.~\ref{alg:schedule}) comprises of three major parts: ordering, slicing of constraints, and planning of atomic tasks.

\paragraph{\textbf{Ordering}} In the first step (line 4), the scheduler computes a partial order ($\prec$) among the given reachability and invariance constraints, to determine which of these constraints must be satisfied before others. In particular, given a pair of constraints, $x_1$ and $x_2$, $x_1 \prec x_2$  if and only if the time of satisfaction of $x_1$ necessarily precede that of $x_2$, based on the time intervals that are assigned to those constraints in $\consexists$ or $\consforall$. 

If one or more of $x_1$ and $x_2$ depends on another symbolic variable, $t$, for satisfaction, then the information in $\tcs'$ is used to determine the presence of a precedence relationship. Consider $t_1$ and $t_3$ from Figure~\ref{fig:tree}; after resolution, the satisfaction of these two constraints depend on the symbolic variable $t$ (as specified in $\tcs'$). Although the value of $t$ is unknown, it can be determined that for any possible value of $t$, $t_3$ will necessarily be satisfied before $t_1$ (i.e., $t_3 \prec t_1$). Overall, for this example, \textsc{ComputeOrder} determines that $t_2 \prec t_3 \prec t_1$. Note that $t_4$ does not appear in this ordering as it needs to be satisfied in parallel with these other constraints.

\paragraph{\textbf{Slicing}} The scheduler then determines the first set of constraints (or tasks) to be carried out based on the order $\prec$ (line 5). In general, the time intervals over these constraint may overlap with each other in an arbitrary way. Recall, however, that each atomic task to be solved by the planner must be in form $F_{[a,b]}(\phi_1) \land G_{[a,b]}(\phi_2)$. Thus, the scheduler must first convert the constraints in $currTasks$ into a set of atomic tasks constraint; this step involves \emph{slicing} one or more constraints in $currTasks$ (line 7). 

\begin{wrapfigure}{r}{0.35\textwidth}
\vspace{-15pt}
  \begin{center}
    \includegraphics[width=0.35\textwidth]{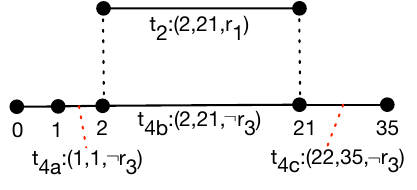}
  \end{center}
  \caption{A slicing example.}
  \label{fig:slicing}
\vspace{-13pt}
\end{wrapfigure}

Due to limited space, we provide the full details of the slicing algorithm in the appendix. We briefly illustrate it here using an example, shown in Figure~\ref{fig:slicing}. For our running example, the first set of constraints to be fulfilled is $\{t_2=(2,21,r_1), t_4=(1,35,\neg r_3)\}$. To generate atomic tasks out of these, the \textsc{Slice} operation slices $t_4$ into three constraints, $t_{4a}, t_{4b}, t_{4c}$; this, in turn, results in the following three atomic tasks:
\begin{align*}
G_{[1,1]}(\neg r_3) \qquad F_{[2,21]}(r_2) \land G_{[2,21]}(\neg r_3) \qquad   G_{[22,35]}(\neg r_3) 
\end{align*}

However, the slicing step may introduce  dependencies among the atomic tasks, especially for the constraints in $\consexists$. For example, suppose that task $F_{[2,21]}(r_2)$ is further split into two slices, $F_{[2,12]}(r_2)$ and $F_{[12,21]}(r_2)$. Since $r_2$  needs to be satisfied only once in  interval $[2,21]$, we would need to ensure that we are not over-constraining the space of possible behaviors by requiring $r_2$ to be satisfied twice, as the slices would imply. To achieve this, we keep track of dependencies among constraints, which are used by the scheduler to remove unnecessary atomic tasks (e.g., remove $F_{[12,21]}(r_2)$ once $r_2$ is satisfied between $[2,12]$). This dependency management is handled inside \textsc{NextTasks} (Algorithm \ref{alg:schedule}, Line 14).

\paragraph{\textbf{Planning atomic tasks}} Once  atomic tasks have been generated through slicing, the scheduler selects the next atomic task and invokes the planner (lines 8-10). If the planner is able to synthesize a plan that satisfies the atomic task, it returns a signal that represents the satisfying trajectory, which is then appended to the cumulative signal $\signal_{plan}$ (line 12); if not, the scheduler terminates by returning the signal that contains a partially satisfactory plan (line 11).

From the synthesized signal, the scheduler extracts the constraints from $\consexists$ and $\consforall$ that were satisfied, along with the concrete time values for their satisfaction (line 13). This information ($satTasks$) is then used to determine the next increment of constraints to be solved.  

Finally, when all of the constraints in $\consexists$ and $\consforall$ have been satisfied, the scheduler terminates by returning $\signal_{plan}$ as the final plan.
\begin{lemma}
Let $s_{plan}$ be the output of $Schedule$($\consexists$, $\consforall$, $TC'$) and $v$ be an assignment of values to symbolic time variables in $TC'$, which are determined during the scheduling step. Then, the following statement holds:
\[
\forall x \in \consexists \cup \consforall \bullet s_{plan} \models Inst(x,v)
\]
\end{lemma}
In other words, the synthesized plan, $s_{plan}$, satisfies all of the reachability and invariance constraints that were generated from the preceding flattening and STR steps.

Building on Lemmas 1, 2, and 3, we finally introduce a theorem to state that our proposed  approach generates a plan that satisfies the given specification $\phi$:
\begin{theorem}
Given specification $\phi$ and $s_{plan}$ as the output of the scheduling algorithm, $s_{plan} \models \phi$.
\end{theorem}

\section{Evaluation}%
\label{sec:evaluation}
This section begins with a detailed description of our experimental setup, including specifications and implementation details, alongside the research questions we aim to address. Following this, we present our findings and conclude with a discussion of our approach over the benchmarks.
\subsection{Experimental Setup}
\begin{table}[!t]
    \centering

\begin{tabular}{|p{0.5cm}|p{7cm}|c|}
\hline
 &  STL Specification & Pattern \\
\hline
 $\phi_1$ & $F_{[0,15]} (R_1) \land F_{[5,25]} (R_2)\land F_{[20,30]} (R_3) \land G_{[0,40]} (\neg O_1)$ & R+A \\
\hline
 $\phi_2$ &  $F_{[0,15]} (R_1 \land F_{[0,15]} (R_2)) \land  G_{[0,40]} (\neg O_1)$ & SV+A\\
\hline
 $\phi_3$ & $F_{[0,15]} (R_1 \land F_{[0,15]} (R_2 \land F_{[0,20]}(R_3 \land F_{[0,15]} (R_1))))$ & SV\\

\hline
 $\phi_4$ &  $F_{[0,15]} G_{[0,10]}(R_1) \land F_{[0,35]}(R_2) \land G_{[0,40]} (\neg O_1)$ & R+A+SB\\
\hline
 $\phi_5$ &  $F_{[0,15]} (R_1 \land F_{[0,20]}  G_{[0,10]} (R_2))$  &SV+SB \\
\hline
\end{tabular}
\vspace{10pt}
\caption{Benchmark STL specifications created from motion planning patterns. Here, R: Reach, A: Avoid, SV: Sequenced Visit, SB: Stabilization.}

\label{tab:specs}

\end{table}

\subsubsection{Specifications.}
We investigated multiple motion planning STL specifications from \cite{DBLP:journals/corr/abs-2305-07766,8859226}. Based on the most common planning patterns, such as Reach (R), Avoid (A), Stabilisation (SB), Sequenced Visits (SV) etc., we created representative STL benchmark specifications as outlined in Table \ref{tab:specs}. These specifications are defined over STL subformulas of the form $R_{i}$ or $O_{i}$ where $R_{i}$ / $O_{i}$  is satisfied if the agent is inside Region $i$ or Obstacle $i$. These subformulas are defined in a similar fashion using conjunction of linear and nonlinear predicates as done in \cite{kurtz2022mixed}. Please refer to \cite{kurtz2022mixed} for more information on how these are defined for rectangular/circular regions.  

\subsubsection{Implementation Details.}We investigate planning from benchmark specifications in two robot exploration environments (similar to Figure \ref{fig:pull}), namely LinEnv and NonLinEnv created using \textsc{stlpy} \cite{kurtz2022mixed}. \textsc{stlpy} has the functionality to encode any arbitrary STL formula, dynamics and actuation limits into constraints and use existing state-of-the-art solvers (Gurobi \cite{SCC.GurobiOptimization2012}, SNOPT \cite{10.1137/S0036144504446096}, etc.,) to generate satisfying plans. Our two environments are both planar but differ in underlying dynamics governing the robot. LinEnv has linear dynamics (Double Integrator) whereas NonLinEnv has nonlinear dynamics (Unicycle). We benchmark our technique against existing MICP methods (for linear dynamics) and other gradient-based techniques like SNOPT (for non-linear dynamics). We use Python to implement our tool\footnote{https://github.com/parvkpr/MCTSTL} while using stlpy and Drake~\cite{drake} to encode the STL constraints. Additionally, we use Gurobi or SNOPT to solve the final constrained optimization problem. All experiments were run on a workstation with an Intel Xeon W-1350 processor and 32 GB RAM. 

\subsubsection{Benchmarks and Research Questions.} We compare against the state-of-the-art techniques proposed in \cite{doi:10.1146/annurev-control-053018-023717} (which we call \emph{standard MICP}) and \cite{kurtz2022mixed} (\emph{reduced MICP}).
Since the standard MICP encoding is only defined for environments with linear dynamics, we compare our technique against reduced MICP encoding for NonLinEnv.  Reduced MICP claims better performance over standard MICP for long horizon and complex specifications due to their efficient encoding of disjunction and conjunction with fewer binary variables. However, standard MICP is faster for short-horizon specifications due to solver-specific presolve routines that leverage the additional binary variables for simplification. 
 
Since our focus is on both short- and long-horizon specification with deep levels of temporal operator nesting, we benchmark against both techniques.  The two main metrics we are concerned with are the time taken for solving and the final robustness values. To make the comparison fair, the total time taken by our technique includes the time taken by the flattener, scheduler, and solvers. 

The two main research questions we investigate in this paper are:
\begin{enumerate}
    \item \textbf{RQ1}: Does our decomposition technique result in shorter solve times?
    \item \textbf{RQ2}: Does our decomposition technique result in higher robustness scores?
\end{enumerate}

\subsection{Results}
 Table \ref{tab:lin} summarizes the results for  \toolname{} performance compared to the baselines. In the tables, N represents the horizon of the specification and D represents the maximum depth of temporal nesting; TO represents a timeout, which means the solver did not terminate despite running it for 30 minutes. In those cases, the solver's output plan robustness is represented as -inf (which means no solution was found in the given time). 
 
 For LinEnv for all the specifications, our robustness values are comparable to the two techniques but our solve times are either lower or comparable to the baselines.
Additionally, for specification $\phi_3$, which has the deepest temporal nesting, our method significantly outperforms both baseline methods that experience timeouts. 

For NonLinEnv for $\phi_1$ and $\phi_4$, the baseline encoding performs better in terms of solving time but \toolname{} only does slightly worse.  However, for specification $\phi_2, \phi_3$ and $\phi_5$, \toolname{} significantly outperforms the baselines.

\subsubsection{Summary.} Our technique excels significantly for nesting depths $>1$ in both LinEnv and NonLinEnv. However, for nesting depths $<1$, the baseline techniques outperform us due to marginal overhead from flattening, scheduling, and solver invocations. The encoding for these specifications involves fewer binary variables, and the preprocessing overhead of using \toolname{} outweighs the performance benefits. Nevertheless, the experiment suggests that our technique is more efficient  for multi-step tasks with deep temporal nesting, outperforming baselines by an order of magnitude.

\begin{table}[!t]
   \centering
    \resizebox{\textwidth}{!}{
    \begin{tabular}{|c|c|c|c|c|c|c|c|c|c|c|c|c|}
        \hline
    {\textbf{Spec}} & \textbf{N} & \textbf{D} &\multicolumn{5}{c|}{\textbf{Solve time (s)}} & \multicolumn{5}{c|}{\textbf{Robustness}} \\
    \hline
        \multicolumn{1}{|c|}{} &  & & \multicolumn{3}{c|}{LinEnv} & \multicolumn{2}{c|}{NonLinEnv} & 
        \multicolumn{3}{c|}{LinEnv} & \multicolumn{2}{c|}{NonLinEnv}\\ 
    \hline
    \multicolumn{1}{|c|}{} &  & &  \cite{doi:10.1146/annurev-control-053018-023717} & \cite{kurtz2022mixed} & \toolname{} & \cite{kurtz2022mixed} & \toolname{}  & \cite{doi:10.1146/annurev-control-053018-023717} & \cite{kurtz2022mixed} & \toolname{} & \cite{kurtz2022mixed} & \toolname{} \\ 
    \hline
    $\phi_1$  & 40 & 0 & \textbf{0.845} & 2.698& 0.891 & \textbf{0.890} & 1.464 & 0.500 & 0.500 & \textbf{0.500} & \textbf{0.430} & 0.572 \\
    \hline
    $\phi_2$ & 30 & 1 & 2.459 & TO & \textbf{0.402} & 12.674 & \textbf{0.892} & 0.491 & -inf & \textbf{0.491} & -inf & \textbf{0.594} \\ 
    \hline
    $\phi_3$ & 60 & 3 &  TO & TO & \textbf{0.874} & 15.829 & \textbf{1.554} & -inf & -inf & \textbf{0.228} & -inf & \textbf{0.065}\\ 
    \hline
    $\phi_4$ & 40 & 2 & \textbf{0.318} & 0.330 & 0.629 & \textbf{1.049} & 1.131 & \textbf{0.494}  & 0.500 & 0.500 & 0.470 & \textbf{0.364} \\ 
    \hline
    $\phi_5$ & 40 & 2 & 2.829 & 28.490 & \textbf{0.694} & 83.193 & \textbf{1.776} &  0.500 & 0.500 & \textbf{0.500} & -inf & \textbf{0.596}  \\ 
    \hline
    \end{tabular}}
    \vspace{10pt}
    \caption{\toolname{} Performance Benchmarking for LinEnv and NonLinEnv against standard MICP (\cite{doi:10.1146/annurev-control-053018-023717})  and reduced MICP (\cite{kurtz2022mixed}).}
    \label{tab:lin}
\end{table}

\section{Related Work}%
\label{sec:related-work}
Trajectory synthesis from STL specifications is an active area of research for which multiple approaches have been proposed in the past few years~\cite{7039363,10160953,8310901,aksaray2016qlearning,Lindemann2019ControlBF}. One of the first papers in this direction involved translating STL specifications into constraints within a Mixed Integer Linear Program (MILP) \cite{7039363}. This approach is sound and complete but faces scalability challenges for long-horizon specifications. To remedy this drawback, the original encoding has been modified by focusing on abstraction-based techniques \cite{Sun2022MultiagentMP} and reducing binary variables via logarithmic encoding \cite{kurtz2022mixed}. Most of these techniques focus on reducing the MILP's complexity to observe performance benefits. 
Recently, the focus has shifted to developing techniques that leverage robustness feedback as a heuristic for trajectory synthesis instead of using MILP. These techniques involve using reinforcement learning \cite{aksaray2016qlearning,kapoor2020model}, search-based techniques \cite{aloor2023follow} and control barrier functions \cite{Lindemann2019ControlBF} to generate STL satisfying trajectories. While these methods offer greater scalability, they are not complete and frequently struggle to accommodate complex specifications because of the intrinsically non-convex optimization problem posed by robustness semantics.

In this work, we focus on MILP-based techniques due to their completeness guarantees and improve their scalability by modifying input STL specifications themselves. However, since we perform structural manipulation of the specifications themselves, our decomposition technique is planner-agnostic and can also use learning- or search-based planners. Decomposition of STL specifications has been studied before in \cite{10.1007/978-3-031-33170-1_14,yu2022model}. However, our work differs from existing work in multiple ways. In \cite{yu2022model}, the authors restrict themselves to an STL fragment that does not allow nesting of temporal operators, while a key contribution of our work is handling deep nested specifications. In \cite{10.1007/978-3-031-33170-1_14}, the authors perform structural manipulation using a tree structure. However, their focus is on multi-agent setups and they handle nested operators conservatively, especially for the eventually operator. This conservative notion generates specification satisfying behavior but it can be overly restrictive. Our interpretation is more flexible and in line with the Boolean semantics defined for the same operators.
\section{Limitations and Future Work}%
\label{sec:conclusion}

In this work, we propose a structural manipulation-based technique for the temporal decomposition of STL specifications, enabling the incremental fulfillment of these specifications. We show our method generates correct-by -construction trajectories that satisfy deeply nested specifications with long time horizons for which existing baseline STL planning techniques struggle. 
 
 While the proposed approach is promising, our current decomposition technique does not handle disjunction. Additionally, the technique is sound but not complete, and designed to prioritize satisfaction over optimality. This limitation stems from incremental planning of objectives, which can be locally optimal compared to global planning, which considers the entire problem space. In future work, we aim to enrich our scheduling algorithm with backtracking capability, which can generate multiple satisfying plans, to overcome this limitation. 
 Furthermore, for probabilistic systems, we plan to employ conformal prediction techniques to overcome compounding modeling error issues. Another avenue for future work is to adapt our theory to accommodate learning and heuristic-based approaches, such as reinforcement learning. Finally, our decomposition theory can potentially be employed for other STL applications, such as falsification, testing and runtime assurance and we plan to investigate that in the future.

\section*{Acknowledgments}
We'd like to thank our reviewers for their insightful feedback. This work was supported in part by the National Science Foundation Award CCF-2144860.
\bibliographystyle{splncs04}
\bibliography{main}
\appendix
\renewcommand{\thesection}{A}
\section{Appendix}%
\label{sec:appendix}
\subsection{Slicing Algorithm and Illustration}

Our slicing algorithm is illustrated in Algorithm~\ref{alg:slicing} and its  application is demonstrated using the running example.  In this example, the first set of constraints to be fulfilled is $\{t_2=(2,21,r_1), t_4=(1,35,\neg r_3)\}$. In the first step (line 3), we identify the lowest and the highest time bounds out of all the constraints in $currTasks$ (which, for the example, would be 1 and 35). Then, in lines 5 to 11,

for the horizon, we identify which constraints are active at each given time step. After this step, in lines 14 to 29,  we first create ``slices'', which involves generating multiple time intervals out of time steps where similar constraints are active. Then, we create constraints of type $\conssym$ and $\consforall$ out of them by combining propositions of constraints of the same type. For the running example, these time slices would be [1,1], [2,21] and [22,35]. 
Finally, the constraints are converted into the following three atomic tasks:
\begin{align*}
G_{[1,1]}(\neg r_3) \qquad F_{[2,21]}(r_2) \land G_{[2,21]}(\neg r_3) \qquad   G_{[22,35]}(\neg r_3) 
\end{align*}
\begin{algorithm}
\caption{SLICE}\label{alg:slicing}
\begin{algorithmic}[1]
\State \textbf{Input}: Set of currTasks $\mathcal{X}^{c}$
\State \textbf{Output}: Set of atomic tasks constraints $\mathcal{X}^{c'}$
\State $t_{min}$, $t_{max}$ := \textsc{Horizon}($currTasks$) 
\State $active := \langle \rangle$
\For{$t$ in $[t_{min}, t_{max}]$} 
\State $active_t := \langle \rangle$
\For{$x=(l, h, p)$ in $\mathcal{X}^{c}$}
\IIf{$t$ in $[l,h]$} $active_t := active_t^\frown \langle x\rangle $\EndIIf 
\EndFor
\State $active := active^\frown active_t$
\EndFor
\State $t_{low} := t_{min}$
\State $\mathcal{X}^{c'}$ = \{\}
\For{$t$ in $[t_{min}+1, t_{max}]$}
\If{$active[t-1] !=active[t]$} 
\State $\mathcal{X}^{temp}$ := $active[t-1]$
\State $\mathcal{X}^{temp}_{\exists}$ := $\{(t_{low}, t-1, \top)\}$
\State $\mathcal{X}^{temp}_{\forall}$ := $\{(t_{low}, t-1, \top)\}$ \For{$x=(l, h, p)$ in $\mathcal{X}^{temp}$}
\If{$x$ in $\consexists$} 
\State $\mathcal{X}^{temp}_{\exists}.p := \mathcal{X}^{temp}_{\exists}.p \land p$ 
\Else{}
\State $\mathcal{X}^{temp}_{\forall}.p := \mathcal{X}^{temp}_{\forall}.p \land p$ 
\EndIf
\EndFor
\State $\mathcal{X}^{c'} := \mathcal{X}^{c'} \cup \mathcal{X}^{temp}_{\exists} \cup \mathcal{X}^{temp}_{\forall} $ 
\State $t_{low} := t$
\EndIf 
\EndFor
\State \Return $\mathcal{X}^{c'}$
\end{algorithmic}
\end{algorithm}
\end{document}